\documentclass[12pt]{iopart}
\usepackage{amssymb}
\newtheorem{prop}{Proposition}
\newcommand{\rn}[1]{\uppercase\expandafter{\romannumeral#1}}
\newcommand{\omax}{\overline{\max}}
\newcommand{\sgn}{{\rm sgn}}
\begin{document}
\title{A New Expression of Soliton Solution to the Ultradiscrete Toda Equation}
\author{Hidetomo Nagai}
\address{Major in Pure and Applied Mathematics, Graduate School of Fundamental Science and Engineering, Waseda University, 3-4-1, Okubo, Shinjuku-ku, Tokyo 169-855, Japan}
\ead{n1a9g8a1i@toki.waseda.jp}
\begin{abstract}
 A new type of multi-soliton solution to the ultradiscrete Toda equation is proposed.  The solution can be transformed into another expression of solution in a perturbation form. A direct proof of the solution is also given.
\end{abstract}
\section{Introduction}
  Ultradiscretization is a limiting procedure to transform a discrete equation with continuous dependent variables into an ultradiscrete equation with discrete dependent variables\cite{tokihiro}.  Ultradiscrete system is applied to the traffic flow\cite{Nishinari1, Nishinari2} and the sorting algorithm\cite{Toda-sort} for example.  An important formula of the procedure is a simple limiting formula,
\begin{equation}  \label{limit}
  \lim_{\epsilon\to+0}\epsilon\log(e^{a/\epsilon}+e^{b/\epsilon})
  = \max(a,b).
\end{equation}
For example, assume a discrete equation,
\begin{equation}
  X_{n+1} = \frac{1+X_n}{X_{n-1}}.
\end{equation}
If we use a transformation of variable $X_n=e^{x_n/\epsilon}$ and take a limit $\epsilon\to+0$, we obtain an ultradiscrete equation,
\begin{equation}
  x_{n+1} = \max(0,x_n)-x_{n-1}.
\end{equation}
If initial values, for example, $x_0$ and $x_1$ are all integer, $x_n$ for any $n$ is also.  The dependent variable $X$ is considered to be discretized in this meaning.\par
  If we apply this procedure to a multi-soliton solution to a discrete soliton equation, we obtain an ultradiscrete solution to an ultradiscrete soliton equation consistently.  For example, a set of equation and solution in an ultradiscrete version is automatically derived by ultradiscretizing the discrete Korteweg--de~Vries equation and its solution where the equation is transformed into a soliton cellular automaton called `box and ball system' with a binary state value\cite{tokihiro,takahashi}.\par
  However, there is a difficulty called `negative problem' in the ultradiscretizing procedure.  If `$+$' in the left side of (\ref{limit}) is replaced by `$-$', the limit by $\epsilon\to+0$ is not well-defined.  Therefore, a multi-soliton solution expressed by a determinant can not be ultradiscretized though that in a perturbation form\footnote{Hirota discovered multi-soliton solutions expressed by 
\begin{equation}
 f= 1+\varepsilon e^{\eta_1} + \varepsilon^2 e^{\eta_2}+\dots+\varepsilon^n e^{\eta_n}
\end{equation}
for arbitrary $\varepsilon$.  In this paper, we call the above form and its ultradiscretized one `perturbation form' referring Ref.\cite{hirota}.} can.  A proof of discrete solution to discrete soliton equation is often realized by using an identity of determinants.  Hence, it is important to find a `determinant` form of ultradiscrete solution corresponding to determinant solutions.  About this problem, Takahashi and Hirota showed a new expression of solution to the ultradiscrete KdV equation recently\cite{uKdV}. The solution is expressed by a form of ultradiscretized permanent which is defined by a signature-free determinant.  It suggests that this type of solution is an ultradiscrete correspondence to a determinant one to a discrete soliton equation.\par
  In this paper, the author proposes a new type of multi-soliton solution to the ultradiscrete Toda equation with a similar structure of expression.  The discrete Toda equation introduced by Hirota\cite{hirota} is
\begin{equation}  \label{discretoda}
\eqalign{
  \log (1+V^m_{n+1})-2\log (1+V^m_n) +\log (1+V^m_{n-1})  \\
  =  \log(1 + \delta^2 V^{m+1}_n)
  - 2\log(1 + \delta^2V^m_n)
  +  \log(1 + \delta^2V^{m-1}_n).
}
\end{equation}
Using transformations including parameters $\epsilon$ and $L$ ($L>0$),
\begin{equation}
  V_n^m=e^{u_n^m/\epsilon}, \qquad \delta=e^{-L/2\epsilon},
\end{equation}
and taking the limit $\epsilon\to+0$, we obtain the ultradiscrete Toda equation\cite{takahashi2,toda}
\begin{equation}
    \fl  u_{n+1}^m -2u_n^m+u_{n-1}^m
  = \max(0, u_n^{m+1}-L) - 2\max(0, u_n^m-L) +\max(0, u_n^{m-1}-L).
\end{equation}
Its bilinear equation can also be ultradiscretized and given by
\begin{equation}  \label{eq-toda}
    f_{n+1}^m +f_{n-1}^m
  = \max(2f_n^m, f_n^{m+1}+f_n^{m-1}-L),
\end{equation}
where $u^m_n$ is related to $f^m_n$ as
\begin{equation}  \label{eq:u-f}
  u_n^m = f_n^{m+1}-2f_n^m+f_n^{m-1}.
\end{equation}
A multi-soliton solution in a perturbation form to (\ref{eq-toda}) is reported in \cite{toda} and is derived by ultradiscretizing that to the discrete Toda equation. It follows a form,
\begin{equation}  \label{pre-toda}
  f^m_n = \max_{\mu_i =0, 1}
    \Bigl(\sum_{i=1}^N\mu_i s_i(m, n)
      + \sum_{1\le i<j\le N}\mu_i \mu_j A_{ij}\Bigr),
\end{equation}
where $\max_{\mu_i=0,1}X(\mu_1,\mu_2,\ldots,\mu_N)$ denotes the maximum value of $X$ in $2^N$ possible cases of $(\mu_1,\mu_2,\ldots,\mu_N)$ replacing each $\mu_i$ by $0$ or $1$.  Detailed information on the right side is shown in Section~\ref{sec:direct proof}.  We refer to the above known type of solution as `type \rn1' in this paper.\par
  We refer to the new type of solution proposed in this paper as `type \rn2'.  The contents of this paper are as follows.  In Section~2, a general form of solution of type \rn2 is proposed and a relation between both types is shown.  In Section~3, a direct proof that the solution of type \rn1 satisfies (\ref{eq-toda}) is given.  Since solution of type \rn2 is equivalent to type \rn1, this proof is also valid for type \rn2.  In Section 4, some conclusions are given.
\section{New type of solution}
 A new type (type \rn2) of multi-soliton solution to the ultradiscrete Toda equation is given by
\begin{equation}  \label{toda-sol1}
  f^m_n = \frac{1}{2}\omax(|s_1(m, n)|,|s_2(m, n)|,\ldots,|s_N(m, n)|),
\end{equation}
where
\begin{equation}
\eqalign{
  &s_i(m, n) = p_i m-\varepsilon_i q_i n +c_i, \\
  &q_i = \min (0, p_i+L) - \min (0, -p_i+L).
}
\end{equation}
Parameters $p_i$ and $c_i$ are real and $\varepsilon_i$ is $1$ or $-1$.  Function $\omax$ is defined by
\begin{equation}
\eqalign{
   & \omax(\varphi_1(m,n),\varphi_2(m,n),\ldots,\varphi_N(m,n)) \\
  =& \max_{\pi_i}(
       \varphi_{\pi_1}(m,n)+\varphi_{\pi_2}(m+1,n-\varepsilon_{\pi_1})
      +\varphi_{\pi_3}(m+2,n-\varepsilon_{\pi_1}-\varepsilon_{\pi_2}) \\
  &\qquad\qquad+\cdots+\varphi_{\pi_N}(m+N-1,n-\varepsilon_{\pi_1}-\varepsilon_{\pi_2}-\cdots-\varepsilon_{\pi_{N-1}}),
}
\end{equation}
where $(\pi_1, \pi_2, \dots , \pi_N)$ denotes an arbitrary permutation of natural numbers $1 \sim N$.  For example, $f^m_n$ in a case of $N=3$ becomes
\begin{eqnarray}
\eqalign{
  \fl f^m_n = \frac{1}{2}
   \max (& |s_1(m,n)| + |s_2(m+1,n-\varepsilon_1)|
         + |s_3(m+2,n-\varepsilon_1-\varepsilon_2)|, \\
         & |s_1(m,n)| + |s_2(m+2,n-\varepsilon_1-\varepsilon_3)|
         + |s_3(m+1,n-\varepsilon_1)|, \\
         & |s_1(m+1,n-\varepsilon_2)| + |s_2(m,n)|
         + |s_3(m+2,n-\varepsilon_1-\varepsilon_2)|, \\
         & |s_1(m+2,n-\varepsilon_2-\varepsilon_3)| + |s_2(m,n)|
         + |s_3(m+1,n-\varepsilon_2)| , \\
         & |s_1(m+1,n-\varepsilon_3)|
         + |s_2(m+2,n-\varepsilon_1-\varepsilon_3)| + |s_3(m,n)|, \\
         & |s_1(m+2,n-\varepsilon_2-\varepsilon_3)|
         + |s_2(m+1,n-\varepsilon_3)| +|s_3(m,n)|).
}
\end{eqnarray}
\par
  Let us transform (\ref{toda-sol1}) to a solution of type \rn1.  Since the parameter $q_i$ depends on $p_i$ as
\begin{equation}  \label{p and q}
  q_i =
   \cases{
     -\min(0, -p_i+L)  &  ($p_i\ge0$) \\
      \min(0,  p_i+L)  &  ($p_i< 0$)
   },
\end{equation}
$q_i$ is odd on $p_i$. Thus, $(p_i, c_i)$ and $(-p_i, -c_i)$ give the same $|s_i(m, n)|$ and we may assume
\begin{equation}  \label{assum}
  p_1 \ge p_2 \ge \dots \ge p_N > 0
\end{equation}
without loss of generality.
Under this assumption, (\ref{toda-sol1}) reduces to
\begin{equation}  \label{trans-formula}
\eqalign{
  f^m_n = \frac{1}{2}
  \max_{\sigma_i= \pm 1}
    & \Bigl(\sum_{i=1}^N \sigma_i s_i
        + \sum_{i=1}^N \Bigl(\frac{1+\sigma_i}{2}
            \Bigl((N-i)p_i + q_i \varepsilon_i
            \sum_{j=i+1}^N \varepsilon_j \Bigr) \\
    & \qquad + \sigma_i \sum_{j=1}^{i-1} \frac{1-\sigma_j}{2}
       (p_i+\varepsilon_i \varepsilon_j q_i)\Bigr) \Bigr),
}
\end{equation}
where $s_i$ denotes $s_i(m, n)$ for abbreviation. The proof of equivalence between (\ref{toda-sol1}) and (\ref{trans-formula}) is shown in Appendix~A.
We obtain from (\ref{trans-formula}),
\begin{equation}  \label{trans-2}
\eqalign{
  f_n^m \Bumpeq
  \frac{1}{2}\max_{\sigma_i=\pm1}\Bigl(& \sum_{i=1}^N(1+\sigma_i)s_i
    + \sum_{i=1}^N \Bigl(\frac{1+\sigma_i}{2}
        \Bigl((N-i)p_i + q_i \varepsilon_i \sum_{j=i+1}^N \varepsilon_j
        \Bigr) \\
  &\qquad + \sigma_i\sum_{j=1}^{i-1}\frac{1-\sigma_j}{2}
    (p_i + \varepsilon_i \varepsilon_j q_i) \Bigr) \Bigr),
}
\end{equation}
where
$f^m_n \Bumpeq g^m_n$ means that $f^m_n$ and $g^m_n$ give the same $u^m_n$ through the right side of transformation (\ref{eq:u-f}).
Moreover, (\ref{trans-2}) reduces to
\begin{equation}
\eqalign{
  \fl f^m_n \Bumpeq
  \max_{\mu_i=0,1}
    \Bigl(
      \sum_{i=1}^N \mu_i
        \Bigl(
          s_i + \frac{(N-i)p_i}{2}
          + q_i \varepsilon_i \sum_{j=i+1}^N \frac{1}{2} \varepsilon_j
          + \sum_{j=1}^{i-1}(p_i + \varepsilon_i \varepsilon_j q_i) \\
          + \frac{1}{2}\sum_{j=i+1}^N
              (p_j + \varepsilon_i \varepsilon_j q_j)
        \Bigr)
      - \sum_{1\le i < j\le N}\mu_j\mu_i
          (p_j + \varepsilon_i \varepsilon_j q_j)
    \Bigr)
}
\end{equation}
using $\mu_i =(1+\sigma_i)/2$.  Hence, exchanging the free parameter $c_i$ as
\begin{equation}
  \fl c_i
  + \frac{(N-i)p_i}{2}
  + q_i \varepsilon_i \sum_{j=i+1}^N \frac{1}{2} \varepsilon_j
  + \sum_{j=1}^{i-1}(p_i + \varepsilon_i \varepsilon_j q_i)
  + \frac{1}{2}\sum_{j=i+1}^N(p_j + \varepsilon_i \varepsilon_j q_j)
  \to c_i,
\end{equation}
we obtain a solution of type \rn1,
\begin{equation}  \label{toda-sol4}
  f^m_n \Bumpeq
  \max_{\mu_i=0,1}
  \Bigl(
    \sum_{i=1}^N \mu_i s_i
    - \sum_{1\le i<j\le N}\mu_i\mu_j
        (p_j + \varepsilon_i \varepsilon_j q_j)
  \Bigr).
\end{equation}
Thus solutions of type \rn1 and \rn2 are equivalent.
\section{Direct Proof of Solution of Type \rn1}  \label{sec:direct proof}
In this section, we prove that a solution of type \rn1 satisfies the ultradiscrete Toda equation.  General form of the solution obeys
\begin{equation}  \label{toda-sol5}
  f^m_n
  = \max_{\mu_i=0,1}
  \Bigl(
    \sum_{i=1}^N \mu_i s_i
    - \sum_{1\le i < j\le N} \mu_i \mu_j A_{ij}
  \Bigr),
\end{equation}
where
\begin{equation}  \label{parameters}
\eqalign{
  & s_i(m,n)=p_im-\varepsilon_iq_in+c_i, \qquad \varepsilon_i=\pm1,\\
  & q_i=\min(0,p_i+L)-\min(0,-p_i+L), \\
  & A_{ij}=\min (\max (a_{ij}, -a_{ji}), \max (-a_{ij}, a_{ji})  ),
    \qquad
    a_{ij} = p_j + \varepsilon_i \varepsilon_j q_j.
}
\end{equation}
In this form, $p_i$ is arbitrary and $f_n^m$ of (\ref{toda-sol5}) reduces to (\ref{toda-sol4}) under the condition (\ref{assum}).\footnote{A mistake exists in the expression of $A_{ij}$ shown in \cite{toda} and is corrected as shown in (\ref{parameters}).}  We give the following proposition about this solution.
\begin{prop}
Solution $f_n^m$ defined by (\ref{toda-sol5}) satisfies a bilinear equation (\ref{eq-toda}), that is,
\begin{equation}  \label{eq-todanew}
  f_{n+1}^m +f_{n-1}^m
  = \max(2f_n^m, f_n^{m+1}+f_n^{m-1}-L).
\end{equation}
\end{prop}
We can assume
\begin{equation}
  |p_1| \ge |p_2|\ge \dots \ge |p_N| > 0
\end{equation}
without loss of generality and then (\ref{toda-sol5}) is expressed by
\begin{equation}  \label{sol:B}
  f^m_n
  = \max_{\mu_i=0,1}
    \Bigl(
      \sum_{i=1}^N \mu_i s_i
      - \sum_{1\le i<j\le N}\mu_i \mu_j \sgn(p_i p_j)|a_{ij}|
    \Bigr)
\end{equation}
where
\begin{equation}
  \sgn(x)=
  \cases{
     1 & ($x>0$) \\
    -1 & ($x<0$)
  }.
\end{equation}
Substituting (\ref{sol:B}) into the left side of (\ref{eq-todanew}), we obtain
\begin{equation}  \label{left side}
\fl\eqalign{
  f_{n+1}^m + f_{n-1}^m
  &= \max_{\mu_i=0,1}
    \Bigl(
      \sum_{1\le i \le N}\mu_i(s_i- \varepsilon_i q_i)
      - \sum_{1\le i< j\le N} \mu_i \mu_j\sgn(p_i p_j)|a_{ij}|
    \Bigr) \\
  &+ \max_{\nu_i=0,1}
    \Bigl(
      \sum_{1\le i \le N}\nu_i(s_i + \varepsilon_i q_i)
      - \sum_{1\le i< j\le N} \nu_i \nu_j \sgn(p_i p_j)|a_{ij}|
    \Bigr).
}
\end{equation}
Introducing new parameters $\lambda_i$ and $\sigma_i$ defined by
\begin{equation}
  \lambda_i=\mu_i+\nu_i,
  \qquad
  \sigma_i=\sgn(p_i)( \mu_i-\nu_i),
\end{equation}
for $1\le i\le N$, we obtain
\begin{equation}  \label{eq:gl}
\eqalign{
  f^m_{n+1} + f^m_{n-1}
  = \max_{(\lambda_i,\sigma_i)}
      \Bigl(
        & \sum_{1\le i\le N} \lambda_i s_i
        - \frac{1}{2}\sum_{1\le i<j \le N} \lambda_i \lambda_j
          \sgn (p_i p_j)|a_{ij}| \\
        &-\sum_{{\scriptstyle 1\le i\le N} \atop {\scriptstyle \lambda_i=1}}
            \sigma_i \varepsilon_i \sgn(p_i) q_i
           -\frac{1}{2}\sum_{{\scriptstyle 1\le i<j \le N}
                \atop{\scriptstyle \lambda_i=\lambda_j=1}}
              \sigma_i \sigma_j|a_{ij}|
      \Bigr).
}
\end{equation}
Note that the pair $(\lambda_i,\sigma_i)$ can be one of the following,
\begin{equation}
  (0, 0),\ (1,1),\ (1,-1),\ (2, 0),
\end{equation}
and $\max_{(\lambda_i,\sigma_i)}X(\lambda_1,\ldots,\lambda_N,\sigma_1,\ldots,\sigma_N)$ denotes the maximum value of $X$ in $4^N$ possible cases of $(\lambda_1,\ldots,\lambda_N,\sigma_1,\ldots,\sigma_N)$ replacing each $(\lambda_i,\sigma_i)$ by one of the above four pairs.  Similarly, terms in the max function of the right side of (\ref{eq-todanew}) become respectively
\begin{eqnarray}
  \fl 2f_n^m
  = \max_{(\lambda_i,\sigma_i)}
    \Bigl(
      \sum_{1\le i\le N} \lambda_i s_i
      - \frac{1}{2}\sum_{1\le i<j \le N}
          \lambda_i \lambda_j \sgn(p_ip_j) |a_{ij}|
      - \frac{1}{2}\sum_{{\scriptstyle 1\le i<j \le N}
            \atop{\scriptstyle \lambda_i=\lambda_j=1}}
          \sigma_i \sigma_j |a_{ij}|
    \Bigr), \label{eq:gr1} \\
\eqalign{
  \fl f_n^{m+1}+f_n^{m-1}-L
  = \max_{(\lambda_i,\sigma_i)}
      \Bigl(
        & \sum_{1\le i\le N} \lambda_i s_i
          -\frac{1}{2}\sum_{1\le i < j \le N}
             \lambda_i \lambda_j \sgn(p_i p_j) |a_{ij}| \\
        & +\sum_{{\scriptstyle 1\le i\le N} \atop {\scriptstyle \lambda_i=1}}
             \sigma_i \sgn(p_i)p_i
          -\frac{1}{2}\sum_{{\scriptstyle 1\le i<j \le N}
              \atop {\scriptstyle \lambda_i=\lambda_j=1}}
            \sigma_i \sigma_j |a_{ij}|
      \Bigr) - L. \label{eq:gr2}
}
\end{eqnarray}
Comparing terms in the max functions of (\ref{eq:gl}), (\ref{eq:gr1}) and (\ref{eq:gr2}), (\ref{eq-todanew}) holds if
\begin{equation}
\eqalign{
  & \max_{\sigma_i=\pm1}
      \Bigl(
         -\sum_{1\le i\le N} \sigma_i \varepsilon_i \sgn(p_i) q_i
         -\frac{1}{2}\sum_{1\le i < j \le N} \sigma_i \sigma_j |a_{ij}|
      \Bigr) \\
= & \max
    \Bigl(
      \max_{\sigma_i=\pm1}
        \Bigl(
          -\frac{1}{2}\sum_{1\le i < j \le N} \sigma_i \sigma_j |a_{ij}|
        \Bigr),\\
  & \qquad\quad
      \max_{\sigma_i=\pm1}
        \Bigl(
          \sum_{1\le i\le N} \sigma_i \sgn(p_i)p_i
          - \frac{1}{2}\sum_{1\le i < j \le N} \sigma_i \sigma_j |a_{ij}|
        \Bigr) - L
    \Bigr).
}
\end{equation}
Considering a symmetry of expression about $p_i$ and $q_i$, we can assume $p_1\ge p_2\ge \dots \ge p_N> 0$.  Under this assumption, $q_i$ is given by
\begin{equation}
  q_i=-\min(0,-p_i+L)
\end{equation}
and $0\le q_i\le p_i$.  Moreover, $a_{ij}=p_j+\varepsilon_i\varepsilon_jq_j\ge0$.  Hence, we give the following proposition to prove.
\begin{prop}
Functions $g_l(\sigma_i)$, $g_{r_1}(\sigma_i)$ and $g_{r_2}(\sigma_i)$ defined by
\begin{equation}
\eqalign{
  g_l(\sigma_i)
  & = -\sum_{i=1}^N \sigma_i\varepsilon_i q_i
      - \frac{1}{2}\sum_{\scriptstyle 1\le i < j \le N}
          \sigma_i \sigma_j a_{ij}, \\
  g_{r_1}(\sigma_i)
  & = -\frac{1}{2}\sum_{\scriptstyle 1\le i<j \le N}
         \sigma_i \sigma_j a_{ij}, \\
  g_{r_2}(\sigma_i)
  & = \sum_{i=1}^N \sigma_i p_i
      - \frac{1}{2}\sum_{\scriptstyle 1\le i<j \le N}
          \sigma_i \sigma_j a_{ij},
}
\end{equation}
satisfy
\begin{equation}  \label{eq-toda3}
  \max_{\sigma_i=\pm 1}g_l(\sigma_i)
  = \max\bigl(
      \max_{\sigma_i=\pm 1} g_{r_1}(\sigma_i),\ 
      \max_{\sigma_i=\pm 1} g_{r_2}(\sigma_i)-L
    \bigr),
\end{equation}
for any $N$ and $p_i$ $(p_1\ge p_2\ge \dots \ge p_N> 0)$.
\end{prop}
\par
  Define maximum values of $g_l(\sigma_i)$, $g_{r_1}(\sigma_i)$ and $g_{r_2}(\sigma_i)$ by $\bar{g}_l$, $\bar{g}_{r_1}$ and $\bar{g}_{r_2}$ respectively.  They are given if $\sigma_i$ satisfies the following conditions,
\begin{equation}
\eqalign{
 \bar{g}_l&:\quad \sigma_i =
    (-1)^i \prod_{l=1}^i\varepsilon_l  \hbox{\quad($i$ : odd)}, \qquad
    (-1)^i \prod_{l=1}^{i-1}\varepsilon_l  \hbox{\quad($i$ : even)}, \\
 \bar{g}_{r_1}&:\quad \sigma_i =
   \prod_{l=1}^i \varepsilon_l \hbox{\quad($i$ : odd)}, \qquad
   -\prod_{l=1}^{i-1} \varepsilon_l \hbox{\quad($i$ : even)}, \\
 \bar{g}_{r_2}&:\quad \sigma_i =
    1 \hbox{\quad($i=1$)}, \qquad
    (-1)^i\varepsilon_1\prod_{l=1}^i\varepsilon_l
     \hbox{\quad($i$ : odd, $i\ge3$)}, \\
  & \qquad (-1)^i\varepsilon_1\prod_{l=1}^{i-1}\varepsilon_l
    \hbox{\quad($i$ : even)}.
}
\end{equation}
Substituting these conditions, we can derive
\begin{eqnarray}
  \fl \bar{g}_l
  &= \sum_{i=1}^N \frac{1+(-1)^i}{4} p_i +q_1
   - \sum_{i=2}^N
       \Bigl(
         \frac{1+(-1)^i}{4} \prod_{l=1}^i \varepsilon_l
         - \frac{1+(-1)^{i-1}}{4}
             \Bigl(
               1+\prod_{l=1}^{i-1} \varepsilon_l
             \Bigr)
       \Bigr) q_i, \label{maxgl} \\
  \fl \bar{g}_{r_1}
  &= \sum_{i=1}^N \frac{1+(-1)^i}{4} p_i
   + \sum_{i=2}^N
       \Bigl(
         \frac{1+(-1)^i}{4} \prod_{l=1}^i \varepsilon_l
         + \frac{1+(-1)^{i-1}}{4}
             \Bigl(
               1-\prod_{l=1}^{i-1} \varepsilon_l
             \Bigr)
       \Bigr) q_i, \label{maxgr1} \\
  \fl \bar{g}_{r_2}
  &= p_1 + \sum_{i=1}^N \frac{1+(-1)^i}{4} p_i
   - \sum_{i=2}^N
       \Bigl(
         \frac{1+(-1)^i}{4} \prod_{l=1}^i \varepsilon_l
         - \frac{1+(-1)^{i-1}}{4}
             \Bigl(
               1+\prod_{l=1}^{i-1} \varepsilon_l
             \Bigr)
       \Bigr) q_i. \label{maxgr2}
\end{eqnarray}
Though the whole proof on the above relations (\ref{maxgl})$\sim$(\ref{maxgr2}) is long and omitted, an important formula for the proof is shown in Appendix~B.\par
  If $p_1\ge L$, then we obtain
\begin{equation}
\eqalign{
  \bar{g}_{r_1} - \bar{g}_l
  &= -q_1
   + \sum_{i=2}^N
       \Bigl(
         \frac{1+(-1)^i}{2} \prod_{l=1}^i \varepsilon_l
         - \frac{1+(-1)^{i-1}}{2} \prod_{l=1}^{i-1} \varepsilon_l
       \Bigr) q_i \\
  &\le -q_1+q_2-q_3 +q_4 -\dots +(-1)^N q_N
  \ \le 0,
}
\end{equation}
and
\begin{equation}
 \bar{g}_{r_2}-\bar{g}_l-L = p_1-q_1-L =0.
\end{equation}
If $p_1< L$, then $q_i=0$ for any $1\le i\le N$ and
\begin{equation}
  \bar{g}_{r_1}-\bar{g}_l = 0,
  \qquad
  \bar{g}_{r_2}-\bar{g}_l-L = p_1-L \le 0.
\end{equation}
Thus a relation
\begin{equation}
  \max(\bar{g}_{r_1}-\bar{g}_l,\ \bar{g}_{r_2}-\bar{g}_l-L) = 0,
\end{equation}
holds and (\ref{eq-toda3}) is satisfied.
\section{Conclusion}
  A new expression of multi-soliton solution to the ultradiscrete Toda equation is proposed.  The solution is equivalent to a solution in a perturbation form.  A direct proof of solution is also given.\par
  The former solution expressed by $\omax$ in (\ref{toda-sol1}) can be derived by ultradiscretization.  For example, let us consider
\begin{equation}
\eqalign{
  F = \sum_{\pi_i}&\psi_{\pi_1}(m,n)\psi_{\pi_2}(m+1,n-\varepsilon_{\pi_1})
        \psi_{\pi_3}(m+2,n-\varepsilon_{\pi_1}-\varepsilon_{\pi_2}) \\
  & \qquad \cdots \psi_{\pi_N}(m+N-1,n-\varepsilon_{\pi_1}
      -\varepsilon_{\pi_2}-\cdots-\varepsilon_{\pi_{N-1}}).
}
\end{equation}
If we use a transformation
\begin{equation}
  \psi_i(m,n)=\cosh(s_i(m,n)/2\varepsilon),
\end{equation}
then we give the right side of (\ref{toda-sol1}) by
\begin{equation}
  \lim_{\epsilon\to+0}\epsilon\log F.
\end{equation}
It suggests that a discrete solution as a correspondence of ultradiscrete solution can exists.  However, such a solution has not yet been found.  Moreover, proof of discrete solution to discrete soliton equation is often realized by using an identity of determinants.  To find the discrete correspondence of solution and to prove an ultradiscrete solution by some ultradiscretized identity are considered to be important future problems.
\ack
 The author is grateful to Prof. Daisuke Takahashi and Prof. Ryogo Hirota for many fruitful discussions and helpful advices.
\appendix
\section{Equivalence between (\ref{toda-sol1}) and (\ref{trans-formula})}
In this appendix, we  prove the following proposition.
\begin{prop}  \label{formula}
\ \\
\begin{equation}  \label{ap:1}
\eqalign{
  & \omax(|s_1(m, n)|,|s_2(m, n)|,\ldots,|s_N(m, n)|) \\
 =& \max_{\sigma_i=\pm1}
      \Bigl(
        \sum_{i=1}^N \sigma_i s_i
        + \sum_{i=1}^N
            \Bigl(
              \frac{1+\sigma_i}{2}
                \Bigl(
                  (N-i)p_i+q_i\varepsilon_i
                   \sum_{j=i+1}^N\varepsilon_j
                \Bigr) \\
  &\qquad     + \sigma_i \sum_{j=1}^{i-1} \frac{1-\sigma_j}{2}
                  (p_i + \varepsilon_i \varepsilon_j q_i)
            \Bigr)
      \Bigr)
}
\end{equation}
where
\begin{eqnarray}
  s_i(m, n) = p_i m - \varepsilon_i q_i n + c_i,
  \qquad
  -1\le \varepsilon_i \le 1, \\
  p_i\ge 0,\qquad q_i\ge 0,
  \qquad
  p_1-q_1\ge p_2-q_2 \ge \dots \ge p_N-q_N\ge 0.  \label{rel3}
\end{eqnarray}
\end{prop}
If $p_i$ and $q_i$ satisfy (\ref{p and q}) and (\ref{assum}), conditions of (\ref{rel3}) is all satisfied.  Therefore, if the above proposition is proved, equivalence between (\ref{toda-sol1}) and (\ref{trans-formula}) is shown.  Note that the key formula shown in \cite{uKdV} is a special case of (\ref{ap:1}) with $\varepsilon_i= \varepsilon_j$ for any $1\le i, j\le N$. \\
 \textbf{Proof.} \
The left side of (\ref{ap:1}) is equal to
\begin{equation}  \label{ap:3}
  \max_{\pi_i}\sum_{i=1}^N
    \Bigl|
      s_{\pi_i}
      + (i-1)p_{\pi_i}
      + q_{\pi_i}\varepsilon_{\pi_i}\sum_{j=1}^{i-1}\varepsilon_{\pi_j}
    \Bigr|,
\end{equation}
where $(\pi_1,\pi_2,\dots,\pi_N)$ is a permutation of natural numbers $1\sim N$.  Noting $|x| = \max(x, -x)$, the above expression is rewritten by
\begin{equation}
\eqalign{
  &\max_{\pi_i}
     \Bigl(
       \max_{\sigma_{\pi_i}=\pm 1}
         \sum_{i=1}^N \sigma_{\pi_i}
           \Bigl(
             s_{\pi_i} + (i-1)p_{\pi_i}
             + q_{\pi_i}\varepsilon_{\pi_i}
                 \sum_{j=1}^{i-1}\varepsilon_{\pi_j}
           \Bigr)
     \Bigr) \\
  =&\max_{\sigma_i =\pm 1}
      \Bigl(
        \sum_{i=1}^N \sigma_i s_i
        + \max_{\pi_i}
            \sum_{i=1}^N \sigma_{\pi_i}
              \Bigl(
                (i-1)p_{\pi_i}
                + q_{\pi_i}\varepsilon_{\pi_i}
                    \sum_{j=1}^{i-1}\varepsilon_{\pi_j}
              \Bigr)
      \Bigr).
}
\end{equation}
Thus, what to prove is the following.
\begin{equation}  \label{ap:6}
\eqalign{
  \fl \max_{\pi_i}
    \sum_{i=1}^N\sigma_{\pi_i}
      \Bigl(
        (i-1)p_{\pi_i}
        + q_{\pi_i}\varepsilon_{\pi_i}
            \sum_{j=1}^{i-1}\varepsilon_{\pi_j}
      \Bigr)  \\
  \fl = \sum_{i=1}^N
    \Bigl(
      \frac{1+\sigma_i}{2}
        \Bigl(
          (N-i)p_i
          + q_i \varepsilon_i \sum_{j=i+1}^N \varepsilon_j
        \Bigr)
      + \sigma_i\sum_{j=1}^{i-1}\frac{1-\sigma_j}{2}
          (p_i + \varepsilon_i \varepsilon_j q_i)
    \Bigr).
}
\end{equation}
We use a mathematical induction on $N$.  The claim is trivial for $N=1$.  Then, let us assume (\ref{ap:6}) holds for a certain $N$ and define $L(\pi_1, \pi_2, \dots, \pi_{N+1})$ by
\begin{equation}
  L(\pi_1, \pi_2, \dots, \pi_{N+1})
  = \sum_{i=1}^{N+1}\sigma_{\pi_i}
      \Bigl(
        (i-1)p_{\pi_i}
        + q_{\pi_i}\varepsilon_{\pi_i}
            \sum_{j=1}^{i-1}\varepsilon_{\pi_j}
      \Bigr).
\end{equation}
For $N+1$, assume a certain permutation $(\pi_1,\pi_2,\dots,\pi_{N+1})$ and let $\alpha$ be a number satisfying $1\le \alpha \le N+1$ and $\pi_{\alpha}=1$.  In the case of $\sigma_1=1$, we have
\begin{equation}
\eqalign{
  L(\pi_1, \dots, \pi_{\alpha-1}, \pi_{\alpha +1},
    \dots, \pi_{N+1}, \pi_{\alpha})
  - L(\pi_1, \pi_2, \dots, \pi_{N+1}) \\
  = (N-\alpha +1)p_1
  + q_1\varepsilon_1\sum_{j=\alpha +1}^{N+1}\varepsilon_{\pi_j}
  - \sum_{i=\alpha +1}^{N+1}\sigma_{\pi_i}
      (p_{\pi_i} +\varepsilon_1\varepsilon_{\pi_i} q_{\pi_i}) \\
  = \sum_{i=\alpha +1}^{N+1}
      (p_1
       - \sigma_{\pi_i}p_{\pi_i}
       + \varepsilon_1\varepsilon_{\pi_i}
           (q_1-\sigma_{\pi_i}q_{\pi_i})
      ) \\
  \ge 0.
}
\end{equation}
In the case of $\sigma_1=-1$, we have
\begin{equation}
\eqalign{
  & L(\pi_{\alpha}, \pi_1, \dots , \pi_{\alpha -1}, \pi_{\alpha +1},
      \dots, \pi_{N+1})
  - L(\pi_1, \pi_2, \dots, \pi_{N+1}) \\
  = &\sum_{i=1}^{\alpha -1}
      (p_1
       + \varepsilon_1 \varepsilon_{\pi_i} q_1
       + \sigma_{\pi_i}
           (p_{\pi_i} + \varepsilon_1 \varepsilon_{\pi_i} q_{\pi_i})
      ) \\
  \ge & 0.
}
\end{equation}
Thus, we obtain
\begin{equation}  \label{indu}
\fl\eqalign{
  \max_{\pi_i}L(\pi_1, \pi_2, \dots , \pi_{N+1}) \\
  = \frac{1+\sigma_1}{2} \max_{\pi_i}L(\pi_1+1, \dots , \pi_N+1, 1)
   + \frac{1-\sigma_1}{2} \max_{\pi_i}L(1, \pi_1+1, \dots , \pi_N+1) \\
  = \frac{1+\sigma_1}{2}
      \Bigl(
        N p_1 +q_1\varepsilon_1\sum_{j=2}^{N+1}\varepsilon_j \\
  \qquad\qquad\qquad + \max_{\pi_i}\sum_{i=1}^N \sigma_{\pi_i+1}
            \Bigl(
              (i-1) p_{\pi_i+1}
              + q_{\pi_i+1}\varepsilon_{\pi_i+1}
                  \sum_{j=1}^{i-1}\varepsilon_{\pi_j+1}
            \Bigr)
      \Bigr)  \\
  + \frac{1-\sigma_1}{2}
      \Bigl(
        \max_{\pi_i}\sum_{i=1}^N \sigma_{\pi_i+1}
          \Bigl(
            (i-1) p_{\pi_i+1}
            + q_{\pi_i+1}\varepsilon_{\pi_i+1}
                \sum_{j=1}^{i-1}\varepsilon_{\pi_j+1}
          \Bigr) \\
        \qquad\qquad\qquad + \sum_{i=1}^N \sigma_{\pi_i+1}(p_{\pi_i+1}
          + \varepsilon_1\varepsilon_{\pi_i+1}q_{\pi_i+1})
      \Bigr)  \\
  = \max_{\pi_i}\sum_{i=1}^N \sigma_{\pi_i+1}
      \Bigl(
        (i-1) p_{\pi_i+1}
        + q_{\pi_i+1}\varepsilon_{\pi_i+1}
            \sum_{j=1}^{i-1}\varepsilon_{\pi_j+1}
      \Bigr) \\
  + \frac{1+\sigma_1}{2}
      \Bigl(
        N p_1
        + q_1 \varepsilon_1 \sum_{j=2}^{N+1} \varepsilon_j
      \Bigr)
  + \frac{1-\sigma_1}{2}
      \sum_{i=1}^N \sigma_{\pi_i+1}
        (p_{\pi_i+1} + \varepsilon_1 \varepsilon_{\pi_i+1} q_{\pi_i+1}).
}
\end{equation}
Using the assumption of the induction, (\ref{indu}) reduces to
\begin{equation}
\fl\eqalign{
  \sum_{i=2}^{N+1}
    \Bigl(
      \frac{1+\sigma_i}{2}
        \Bigl(
          (N+1-i)p_i + q_i \varepsilon_i \sum_{j=i+1}^N \varepsilon_j
        \Bigr)
      + \sigma_i\sum_{j=2}^{i-1}\frac{1-\sigma_j}{2}
          \Bigl(
            p_i + \varepsilon_i \varepsilon_j q_i
          \Bigr)
    \Bigr) \\
  \qquad \qquad \qquad \qquad
  + \frac{1+\sigma_1}{2}
      \Bigl(
        N p_1 +q_1 \varepsilon_1 \sum_{j=2}^{N+1} \varepsilon_j
      \Bigr)
  + \frac{1-\sigma_1}{2}
      \Bigl(
        \sum_{i=2}^{N+1} \sigma_i(p_i + \varepsilon_1 \varepsilon_i q_i)
      \Bigr) \\
  = \sum_{i=1}^{N+1}
      \Bigl(
        \frac{1+\sigma_i}{2}
          \Bigl(
            (N-i)p_i + q_i \varepsilon_i \sum_{j=i+1}^N \varepsilon_j
          \Bigr)
        + \sigma_i \sum_{j=1}^{i-1} \frac{1-\sigma_j}{2}
            \Bigl(
              p_i + \varepsilon_i \varepsilon_j q_i
            \Bigr)
      \Bigr).
}
\end{equation}
This is equal to the right side of (\ref{ap:6}) for $N+1$ and the induction holds.  Thus this completes the proof of Prop.\ref{formula}.\hfill $\Box$
\section{Proposition for proof on (\ref{maxgl})$\sim$(\ref{maxgr2})}
  To prove (\ref{maxgl})$\sim$(\ref{maxgr2}), the following proposition is important.  The proof on the proposition is shown in this appendix.
\begin{prop}  \label{appprop} \ \\
  Define $g(\sigma_1, \sigma_2, \dots , \sigma_N)$ by
\begin{equation}
  g(\sigma_1, \sigma_2, \dots , \sigma_N)
  = -\sum_{1\le i<j\le N} \sigma_i \sigma_j p_j,
\end{equation}
where $N$ is a natural number, $\sigma_i$ is $+1$ or $-1$ and
\begin{equation}
  p_1\ge p_2\ge \dots \ge p_N> 0.
\end{equation}
Then,
\begin{equation}  \label{prop:app}
  g(\sigma_1, \sigma_2, \dots, \sigma_{N-n-1}, \bar{\sigma}_{N-n},
    \bar{\sigma}_{N-n+1}, \dots, \bar{\sigma}_N)
  \ge
  g(\sigma_1, \sigma_2, \dots, \sigma_N)
\end{equation}
for any $0\le n\le N-2$, where $\bar{\sigma}_i$ is given by
\begin{equation}  \label{pro:sgm}
 \bar{\sigma}_i =
  \cases{
    1 & ($\sum_{l=1}^{N-n-1}\sigma_l
         +\sum_{l=N-n}^{i-1}\bar{\sigma}_l\le 0$) \\
  - 1 & ($\sum_{l=1}^{N-n-1}\sigma_l
         +\sum_{l=N-n}^{i-1}\bar{\sigma}_l>0$)
  }
\end{equation}
\end{prop}
\textbf{Proof.} \
  We use a mathematical induction on $n$.  For $n=0$, we have
\begin{equation}
    g(\sigma_1 , \dots , \sigma_{N-1}, \bar{\sigma}_N)
  - g(\sigma_1, \dots , \sigma_{N-1}, -\bar{\sigma}_N)
  = -2\bar{\sigma}_Np_N\sum_{i=1}^{N-1}\sigma_i
  \ge 0,
\end{equation}
and (\ref{prop:app}) holds.  Assume (\ref{prop:app}) for a certain $n=k$ $(0\le k\le N-2)$, that is,
\begin{equation}
  g(\sigma_1, \dots ,\sigma_{N-k-1}, \bar{\sigma}_{N-k},
    \bar{\sigma}_{N-k+1} , \dots , \bar{\sigma}_N)
  \ge
  g(\sigma_1, \dots , \sigma_N).
\end{equation}
Then, what to prove for induction in the case of $n=k+1$ is
\begin{equation}  \label{app:subs}
\eqalign{
  g(\sigma_1, \dots, \sigma_{N-k-2}, \bar{\sigma}_{N-k-1},
      \bar{\sigma}_{N-k}, \dots,  \bar{\sigma}_N) \\
  \qquad - g(\sigma_1, \dots , \sigma_{N-k-2}, -\bar{\sigma}_{N-k-1},
      \hat{\sigma}_{N-k}, \dots,  \hat{\sigma}_N)
  \ge 0,
}
\end{equation}
where
\begin{equation}
\eqalign{
 \bar{\sigma}_{N-k-1} =
  \cases{
   1 & ($\sum_{l=1}^{N-k-2}\sigma_l\le 0$) \\
  -1 & ($\sum_{l=1}^{N-k-2}\sigma_l>0$)
  }, \\
 \bar{\sigma}_i =
  \cases{
   1 & ($\sum_{l=1}^{N-k-2}\sigma_l + \bar{\sigma}_{N-k-1}
        + \sum_{l=N-k}^{i-1}\bar{\sigma}_l\le 0$) \\
  -1 & ($\sum_{l=1}^{N-k-2}\sigma_l + \bar{\sigma}_{N-k-1}
        + \sum_{l=N-k}^{i-1}\bar{\sigma}_l>0$)
  }, \\
 \hat{\sigma}_i =
  \cases{
   1 & ($\sum_{l=1}^{N-k-2}\sigma_l - \bar{\sigma}_{N-k-1}+ \sum_{l=N-k}^{i-1}\hat{\sigma}_l\le 0$) \\
  -1 & ($\sum_{l=1}^{N-k-2}\sigma_l - \bar{\sigma}_{N-k-1}+\sum_{l=N-k}^{i-1}\hat{\sigma}_l>0$)
  }.
}
\end{equation}
The left side of (\ref{app:subs}) reduces to
\begin{equation}  \label{g:subs}
\eqalign{
  \fl -\Bigl(
   2\sum_{i=1}^{N-k-2}\sigma_i\bar{\sigma}_{N-k-1}p_{N-k-1}
   + \sum_{i=1}^{N-k-2} \sum_{j=N-k}^N \sigma_i
       (\bar{\sigma}_j - \hat{\sigma}_j) p_j \\
   + \bar{\sigma}_{N-k-1} \sum_{j=N-k}^N
       (\bar{\sigma}_j + \hat{\sigma}_j) p_j
   + \sum_{i=N-k}^N \sum_{j=i+1}^N
       (\bar{\sigma}_i \bar{\sigma}_j
        - \hat{\sigma}_i \hat{\sigma}_j) p_j
  \Bigr).
}
\end{equation}
Let us define $S$ by $\sum_{i=1}^{N-k-2}\sigma_i$ and consider two cases of $S>0$ or $S\le 0$.\par
  In the case of $S>0$, $\bar{\sigma}_i=-1$ for $N-k-1\le i\le N-k+S-1$ since $S + \sum_{l=N-k-1}^{i-1}\bar{\sigma}_l>0$, and $\hat{\sigma}_i=-1$ for $N-k \le i\le N-k+S+1$.  Therefore $\bar{\sigma}_i$ and $\hat{\sigma}_i$ are given by
\begin{equation}
\eqalign{
     (\bar{\sigma}_{N-k-1}, \bar{\sigma}_{N-k}, \dots, \bar{\sigma}_N)
  &= (\underbrace{-1, -1, \dots, -1}_S, 1, -1, 1, -1, 1, \dots), \\
     (-\bar{\sigma}_{N-k-1}, \hat{\sigma}_{N-k}, \dots, \hat{\sigma}_N)
  &= (1, \underbrace{-1, -1, \dots, -1}_S, -1, 1, -1, 1, \dots).
}
\end{equation}
Hence,
\begin{equation}
  \hat{\sigma}_i =
  \cases{
   -\bar{\sigma}_i & ($i = N-k+S-1$) \\
   \bar{\sigma}_i & (otherwise)
  },
\end{equation}
and (\ref{g:subs}) reduces to
\begin{equation}
\fl\eqalign{
  &-\Bigl(
  -2Sp_{N-k-1}  +2Sp_{N-k+S-1}
  -2\sum_{j=N-k}^{N-k+S-2} \bar{\sigma}_j p_j
  -2\sum_{j=N-k+S}^N \bar{\sigma}_j p_j \\
  &\qquad\qquad + 2\sum_{j=N-k+S}^N \bar{\sigma}_j p_j
  +2p_{N-k+S-1} \sum_{i=N-k}^{N-k+S-2} \bar{\sigma}_i
  \Bigr) \\
  =& 2\Bigl(
    Sp_{N-k-1}  -Sp_{N-k+S-1}
    -\sum_{j=N-k}^{N-k+S-2}p_j
    +(S-1)p_{N-k+S-1}
  \Bigr) \\
  \ge&  2\Bigl(
    Sp_{N-k-1}  -Sp_{N-k+S-1}
    -(S-1)p_{N-k}
    +(S-1)p_{N-k+S-1}
  \Bigr) \\
  =& 2(p_{N-k} -p_{N-k+S-1}) \ge0  .
}
\end{equation}
Similarly, in the case of $S\le 0$, $\bar{\sigma}_i$ and $\hat{\sigma}_i$ are given by
\begin{equation}
\eqalign{
  (\bar{\sigma}_{N-k-1}, \bar{\sigma}_{N-k}, \dots, \bar{\sigma}_N)
  &= (\underbrace{1,  1, \dots, 1}_{-S}, 1, -1, 1, -1, 1, \dots), \\
  (-\bar{\sigma}_{N-k-1}, \hat{\sigma}_{N-k}, \dots, \hat{\sigma}_N)
  &= (-1, \underbrace{1, 1, \dots, 1}_{-S}, 1, 1, -1, 1, \dots).
}
\end{equation}
Hence,
\begin{equation}
  \hat{\sigma}_i =
  \cases{
    -\bar{\sigma}_i & ($i = N-k-S$) \\
     \bar{\sigma}_i & (otherwise)
  },
\end{equation}
and (\ref{g:subs}) reduces to
\begin{equation}
\eqalign{
  & -\Bigl(
    2Sp_{N-k-1}  -2Sp_{N-k-S}
    + 2\sum_{j=N-k}^{N-k-S-1} \bar{\sigma}_j p_j
    + 2\sum_{j=N-k-S+1}^N \bar{\sigma}_j p_j \\
  & \qquad\qquad - 2\sum_{j=N-k-S+1}^N \bar{\sigma}_j p_j
    - 2\sum_{i=N-k}^{N-k-S-1} \bar{\sigma}_i p_{N-k-S}
  \Bigr) \\
 =& 2\Bigl(
    -Sp_{N-k-1} + Sp_{N-k-S}
    -\sum_{j=N-k}^{N-k-S-1}p_j -Sp_{N-k-S}
    \Bigr) \\
\ge& 2S\Bigl(-p_{N-k-1}+p_{N-k-S}+p_{N-k}-p_{N-k-S} \Bigr)\ge0.
}
\end{equation}
As a result, (\ref{app:subs}) holds and this completes the proof of Prop.\ref{appprop}.\hfill $\Box$
\section*{References}

\end{document}